\documentclass[acmtog,screen,nonacm]{acmart}
\AtBeginDocument{%
  }

\usepackage{mathtools}
\usepackage{algorithmic}
\usepackage{graphicx}
\usepackage{accents}
\usepackage{textcomp}
\usepackage{multirow}
\usepackage{epstopdf}
\usepackage{listings}

\usepackage{listings}
\usepackage{xcolor}

\usepackage{hyperref}
\usepackage{orcidlink}
 
\definecolor{codegreen}{rgb}{0,0.6,0}
\definecolor{codegray}{rgb}{0.5,0.5,0.5}
\definecolor{codepurple}{rgb}{0.58,0,0.82}
\definecolor{codeblue}{rgb}{0.0,0.4,0.8}
\definecolor{backcolour}{rgb}{0.95,0.95,0.92}
 
\lstdefinestyle{mystyle}{
    backgroundcolor=\color{backcolour},   
    commentstyle=\color{codegreen},
    keywordstyle=\bfseries\color{codeblue},%
    numberstyle=\tiny\color{codegray},
    stringstyle=\color{codepurple},
    basicstyle=\ttfamily\footnotesize,
    breakatwhitespace=false,         
    breaklines=true,                 
    captionpos=t,                    
    keepspaces=true,                 
    numbers=none,%
    xleftmargin=1.5em,
    framexleftmargin=1.5em,
    numbersep=5pt,                  
    showspaces=false,                
    showstringspaces=false,
    showtabs=false,                  
    tabsize=2
}

\usepackage{booktabs}
\usepackage[caption=false,font=footnotesize]{subfig}
\usepackage{bm}

\newcommand{\mb}{\mathbf}
\newcommand{\mc}{\mathcal}
\newcommand{\mbb}{\mathbb}

\newcommand{\dd}{\mathop{}\!\mathrm{d}}
\newcommand{\FOV}{\ensuremath{\mc{X}}}

\DeclareMathOperator*{\argmin}{\arg\min}

\DeclareMathOperator{\sign}{sign}
\DeclareMathOperator{\Var}{Var}

\usepackage{pgfplots}
\usepackage{tikz-3dplot}
\pgfplotsset{compat=1.16}
\usetikzlibrary{patterns,external}
\usepgfplotslibrary{colormaps}

\usepackage{tabularx}
\usepackage{amsthm}
\theoremstyle{definition}

\newtheorem{theorem}{Theorem}
\newtheorem{corollary}[theorem]{Corollary}
\newtheorem{proposition}[theorem]{Proposition}

\author{Dongwoon Hyun\textsuperscript{*}, Samuel Beuret\textsuperscript{*}, \MakeLowercase{and} Jeremy J. Dahl\textsuperscript{*}\\
\NoCaseChange{\normalfont\textsuperscript{*}Stanford University, Stanford, CA, USA }}

\copyrightyear{2026}
\begin{document}

\title{The Contrast Order: An Order-Based Image Quality Criterion for Nonlinear Beamformers}

\thanks{This research was supported in part by the National Institute of Biomedical Imaging and Bioengineering under Grant K99-EB032230. Dongwoon Hyun\orcidlink{0000-0003-2625-8109} was with Stanford University when this work was started. He is now with Aurora. Samuel Beuret\orcidlink{0000-0001-6558-0892} (email: sbeuret@stanford.edu) and Jeremy J. Dahl\orcidlink{0000-0001-9877-452X} (email: jjdahl@stanford.edu) are with the Department of Radiology, School of Medicine, Stanford University, Stanford, CA, USA}

\begin{abstract}
\noindent\rule{\linewidth}{0.4pt}
    Many modern ultrasound beamformers report improved image quality when evaluated using classical criteria like the contrast ratio and contrast-to-noise ratio, which are based on summary statistics of regions of interest (ROIs). However, nonlinear beamformers and post-processing methods can substantially alter these statistics, raising concerns that the reported improvements may reflect changes in dynamic range or remapping rather than a reflection of true information gain, such as clutter suppression. Introduced recently, the generalized contrast-to-noise ratio (gCNR) is designed to address these concerns, but it requires implementation-dependent density estimation. We propose herein a new, simpler, image quality criterion, called the \emph{contrast order} (CO), defined as the expected value of the sign of the difference in brightness between two ROIs. The CO is invariant under all strictly monotonic transformations of the image values, as it depends only on their relative ordering, and is interpretable as the probability that one ROI is brighter than the other minus the probability that it is darker. Unlike the gCNR, the CO has a simple unbiased estimator whose variance provably decreases with the number of samples in each ROI.
  We further propose the \emph{effective contrast ratio} (ECR), which calibrates the contrast order to the familiar contrast ratio such that the two coincide under ideal Rayleigh-speckle statistics. Together, the CO and ECR provide order- and sign-preserving, dynamic-range-invariant criteria for evaluating lesion contrast, offering a principled alternative to classical and newer image quality criteria when assessing modern beamformers.
  \noindent\rule{\linewidth}{0.4pt}
\end{abstract}

\authorsaddresses{}
\maketitle
\renewcommand{\shortauthors}{Hyun, Beuret, and Dahl}

\section{Introduction}
Defining and measuring image quality in a precise and actionable manner is a major challenge in medical ultrasound imaging, particularly when attempting to link image quality directly to diagnostic outcomes. Clinician reader studies are often helpful to evaluate new imaging approaches in diagnostic outcomes, but they are often non-quantitative estimates of image quality that are subject to inter-observer variability, differences in training, and individual preferences, and often require the imaging approach to be well beyond the developmental stage. As a practical alternative to reader studies, a variety of quantitative image quality criteria have been developed to characterize contrast, resolution, and noise properties of ultrasound images. These criteria are commonly used as surrogates for lesion detectability and overall diagnostic performance. However, many of the image quality metrics in widespread use today were designed under specific statistical and signal-processing assumptions that are increasingly violated by modern imaging methods.

Classical image quality criteria such as the contrast ratio (CR), contrast-to-noise ratio (CNR) \cite{patterson1983improvement}, signal-to-noise ratio (SNR), and point-spread-function (PSF)-based resolution metrics were largely developed for images produced by linear delay-and-sum (DAS) beamforming followed by the well-characterized transformation of envelope detection. Under these conditions, image values retain a direct and interpretable relationship to underlying echo amplitudes, and speckle statistics are well approximated by Rayleigh distributions. Within this regime, these metrics admit clear interpretations. For example, the CNR has been shown to describe lesion detectability for an ideal observer under Rayleigh scattering assumptions \cite{smith1983low,zemp2005detection,abbey2006observer,abbey2010optimal,nguyen2016task,hyun2021ultrasound}, while speckle autocorrelation has been linked to system resolution \cite{wagner1988fundamental}.

Contemporary ultrasound imaging research increasingly blurs the traditional boundary between beamforming and image processing. While beamforming was historically implemented as a linear reconstruction of reflectivity from radiofrequency channel data (e.g., DAS), modern systems frequently incorporate nonlinear operations at multiple stages of the processing pipeline. These nonlinearities arise both intentionally, through adaptive \cite{synnevag2009benefits,asl2010eigenspace} and coherence-based \cite{matrone2014delay,mallart1994adaptive,li2003adaptive,camacho2009phase,lediju2011short} beamformers designed to suppress clutter or emphasize reliable signals, and unavoidably, through post-processing operations such as dynamic range compression, contrast enhancement, and display mapping.

As demonstrated by Hverven et al. \cite{hverven2017influence} and Rindal et al. \cite{rindal2019effect}, these nonlinear operations can substantially alter the statistical distributions of image values, even when the underlying information content of the image remains unchanged. Consequently, improvements reported in traditional image quality metrics may reflect changes in image statistics rather than genuine gains in lesion detectability or clutter suppression. This statistical mismatch raises a fundamental concern: when image quality metrics are sensitive to arbitrary nonlinear transformations of image values, they risk conflating cosmetic changes in appearance with meaningful improvements in information content.

One historical response to this challenge has been to constrain comparisons to beamformers that preserve linearity, or to regard nonlinear post-processing as ``merely cosmetic'' \cite{smith1983low}. From this perspective, comparisons between linear and nonlinear methods have often been considered inherently unfair unless all methods obey the same statistical assumptions, particularly those governing speckle statistics and image amplitude distributions. This viewpoint substantially restricts the class of beamformers and image reconstruction algorithms that can be evaluated, excluding many modern approaches whose primary objective is not to preserve traditional speckle statistics, but rather to improve the separability of clinically relevant structures through adaptive, coherence-based, or otherwise nonlinear operations.

An alternative and increasingly influential approach is to adopt information-theoretic criteria. Rodriguez-Molares et al. \cite{rodriguez2019generalized} made the first explicit effort towards evaluating nonlinear beamformers with the generalized CNR (gCNR) (prior to this, Nguyen et al. \cite{nguyen2016task} proposed the Kullback-Leibler divergence, which also achieves the same goal \cite{hyun2021ultrasound}, although the authors focused on Rayleigh scattering at the time). The gCNR describes the fundamental separability of two regions of interest (ROIs) by an information-theoretic ideal observer, and is equal to one minus its error rate \cite{hyun2021ultrasound}. The gCNR, also known as the \emph{total variation distance}, is invariant under all injective dynamic range transformations (DRTs) because it depends only on the probability distributions of the ROIs and not the image values themselves, making it an excellent choice for comparing nonlinear beamformers (we have also proposed a similar DRT-invariant spatial resolution criterion based on self-mutual information \cite{hyun2021info}).

However, the gCNR corresponds to a particularly strong notion of DRT invariance. Image quality criteria are distinguished by the class of DRTs under which they are invariant. Different DRT invariance classes induce different equivalence relations between images and therefore give rise to different notions of image quality. The gCNR's invariance to all injective DRTs produces a coarse equivalence relation that discards all mathematical structure on the image values, including ordering (see Sec.~\ref{sec:evaluating} for a formal discussion).

Furthermore, implementing the gCNR requires density estimation to approximate the probability distributions of image intensities within ROIs \cite{sheather2004density}. There are, however, no unique way to perform density estimation and the gCNR computation is therefore implementation-dependent. If implemented naively, the gCNR can be influenced by the choice of parameters in ways that are difficult to quantify \cite{hyun2021ultrasound, schlunk2023methods} even though density estimation can be systematized to avoid any bias in the estimation of the gCNR \cite{arnestad2024robust}.

Nonetheless, the potential complexity of density estimation led us to wonder whether there exists simpler alternatives to gCNR to robustly measure lesion contrast.
Therefore, our goal in this paper is to propose two such metrics and, \textit{in fine}, provide a principled framework for evaluating modern nonlinear beamformers. Instead of injective DRTs, we consider an invariance class of \emph{strictly monotonic} DRTs, leading to an equivalence relation that preserves relative ordering of image brightness while remaining insensitive to nonlinear remappings commonly applied for visualization and display. This level of invariance is obeyed by the two new criteria we introduce : the \textbf{contrast order} (CO) and its speckle-calibrated equivalent, the \textbf{effective contrast ratio} (ECR).

The CO provides a signed, order-based measure of contrast with an invariance property equivalent in practice to that of gCNR for DRTs commonly used in ultrasound imaging. Importantly, the CO has a simple unbiased estimator based on finite ROI samples with variance governed by the number of independent samples drawn from each ROI. The ECR further calibrates the CO against Rayleigh-speckle statistics to yield a familiar ratio-based interpretation.

The remainder of this paper is organized as follows. Section~\ref{sec:evaluating} reviews existing image quality criteria and formalizes the effects of DRTs. Section~\ref{sec:contrastorder} introduces the CO and ECR and analyzes their mathematical properties. Section~\ref{sec:examples} demonstrates their behavior through simulated and experimental examples. Finally, Section~\ref{sec:discussion} discusses implications, limitations, and directions for future work.

\section{Evaluating Ultrasound Image Quality}
\label{sec:evaluating}
Here, we provide definitions that will allow us to precisely state the invariance properties required of modern image quality criteria.

\subsection{Image Statistics}
\begin{figure}[htb]
  \centering
  \subfloat[Field of View $\mc{X}\subset\mathbb{R}^d$]{
    \tdplotsetmaincoords{70}{110}
    \begin{tikzpicture}[tdplot_main_coords, scale=1.2]
      \draw[thick,->,>=stealth] (0,0,0) -- (2,0,0);
      \draw[thick,->,>=stealth] (0,0,0) -- (0,2,0);
      \draw[thick,->,>=stealth] (0,0,0) -- (0,0,-3);
      \draw[thick,dashed, color=black, fill=red, fill opacity=0.2] (0,0.25,-0.25) -- (0,0.25,-2.75) -- (0,1.75,-2.75) -- (0,1.75,-0.25) -- (0,0.25,-0.25);
      \draw[] (0,1.75,-2.75) node[anchor=south east]{$\mc{X}$};
    \end{tikzpicture}
  }
  \hfill
  \subfloat[Image $\phi:\mc{X}\rightarrow\mc{A}$]{
    \makebox[.25\textwidth]{\includegraphics[width=.15\textwidth]{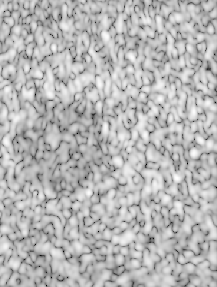}}
  }
  \\
  \subfloat[Histogram $f$ of $\phi$ on the alphabet $\mc{A}$]{
    \begin{tikzpicture}
      \begin{axis}[
          width = 0.4\textwidth,
          height = 0.2\textwidth,
          axis lines = middle,
          axis line style = thick,
          xmin = -.2,
          xmax = 4.2,
          ymin = -.12,
          ymax = .9,
          ticks = none,
          x label style={anchor=north},
          y label style={anchor=west},
          xlabel = $\mc{A}$,
          ylabel = $\textrm{Pr}(a)$,
        ]
        \def\ss{1}
        \addplot[domain=0:4,samples=100,color=black,thick]{x/\ss*exp(-(x)^2/(2*\ss))};
        \shade[
          left color={rgb,1:red,.1;green,.1;blue,.1},
          right color={rgb,1:red,1;green,1;blue,1},
          middle color={rgb,1:red,.7;green,.7;blue,.7}]
        (0,-.02) -- (0,-.1) -- (4,-.1) -- (4,-.02) -- cycle;
        \draw(0,-.02) -- (0,-.1) -- (4,-.1) -- (4,-.02) -- cycle;
        \node at (axis cs:6.5,3.6) {$f$};
      \end{axis}
    \end{tikzpicture}
  }
  \\
  \subfloat[Image space]{
    \begin{tikzpicture}
      \begin{axis}[
          width = 0.48\textwidth,
          height = 0.23\textwidth,
          axis lines = middle,
          axis line style = thick,
          xmin = -2.2,
          xmax = 2.2,
          ymin = -.12,
          ymax = .9,
          ticks = none,
          x label style={anchor=north},
          y label style={anchor=west},
          xlabel = $\mc{A}_\textrm{log}$,
          ylabel = $\textrm{Pr}(a)$,
        ]
        \def\ss{1}
        \addplot[domain=-2.2:2,samples=100,color=black,thick]{exp(x)^2/\ss*exp(-(exp(x))^2/(2*\ss))};
        \shade[
          left color={rgb,1:red,.1;green,.1;blue,.1},
          right color={rgb,1:red,1;green,1;blue,1},
          middle color={rgb,1:red,.7;green,.7;blue,.7}]
        (-2,-.02) -- (-2,-.1) -- (1,-.1) -- (1,-.02) -- cycle;
        \draw(-2,-.02) -- (-2,-.1) -- (2,-.1) -- (2,-.02) -- cycle;
        \node at (axis cs:6.5,3.6) {$f$};
      \end{axis}
    \end{tikzpicture}
  }
  \caption{(a) A 2D FOV $\mc{X}$ visualized in 3D space. (b) An image $\phi$ assigns values in $\mc{A}$ to each coordinate in $\mc{X}$. (c) The histogram of $\phi$ gives the relative occurrence of each image value in $\mc{A}$.
  }
  \label{fig:bmode}
\end{figure}

Mathematically, we can interpret an image as a function $\phi:\FOV\rightarrow\mc{A}$ mapping each point $x$ of some \emph{field of view} (FOV) $\FOV$---representing the set of all positions where the image is defined---into an \emph{alphabet} $\mc{A}\subset\mathbb{R}$, i.e. the range of all possible image values. In the following, we will generally consider images prior to compression where $\mc{A}=[0, \infty)$, i.e. the alphabet, is the set of all non-negative real numbers, unless mentioned otherwise.

Furthermore, we can associate a probability density function (PDF) $f: \mc{A}\rightarrow[0, \infty)$ to the image, denoted herein as its \emph{histogram}. The histogram quantifies how the output values of $\phi$ are distributed. Mathematically, it follows the usual definition of a PDF, namely that
\begin{equation}
    \textrm{Pr}[\alpha \leq a < \beta] = \int_{\alpha}^{\beta} f (a) \dd a,\ \forall a\in \mc{A},
    \label{eq:PDF_def}
\end{equation}
where $\alpha\in\mc{A}$, $\beta\in\mc{A}$, $\alpha<\beta$ are two arbitrary bounds and $\textrm{Pr}[\alpha \leq a < \beta]$ is the probability that an image value $a=\phi(x)$ falls between these bounds. Figure~\ref{fig:bmode} illustrates these definitions.

In this work, we are interested in summarizing the distribution of image values as a single statistical descriptor. To do so, we extensively rely on the expected value, defined as  
\begin{align}
  \mbb{E}[h(\phi)] = \int_{\mc{A}} h(a) \,f(a)\dd a.
  \label{eq:EA_Alphabet}
\end{align}
In \eqref{eq:EA_Alphabet}, $h$ is an arbitrary function and $h(\phi)$ denotes the image obtained by applying $h$ to every value $\phi(x)$ in the field of view $\FOV \ni x$.

\subsection{Selecting Regions of Interest}
\begin{figure}[htb]
  \centering
  \subfloat[ROIs $\mc{X}_A,\mc{X}_B$]
  {
    \tdplotsetmaincoords{70}{110}
    \begin{tikzpicture}[tdplot_main_coords]
      \draw[->,>=stealth] (0,0,0) -- (2,0,0);
      \draw[->,>=stealth] (0,0,0) -- (0,2,0);
      \draw[->,>=stealth] (0,0,0) -- (0,0,-3);
      \draw[thick,dashed, color=black] (0,0.25,-0.25) -- (0,0.25,-2.75) -- (0,1.75,-2.75) -- (0,1.75,-0.25) -- (0,0.25,-0.25);
      \draw[] (0,1.75,-2.75) node[anchor=south east]{$\mc{X}$};
      \tikzset{zyplane/.style={canvas is zy plane at x=#1,very thin}}
      \begin{scope}[zyplane=0]
        \filldraw[thick, dashed, color=black, fill=green, fill opacity=0.2, very thick](-1.5,.6) circle (.3);
        \filldraw[thick, dashed, color=black, fill=blue, fill opacity=0.2, very thick](-1.5,1.4) circle (.3);
        \draw[] (-1.5,0.6,0) node{$\mc{X}_A$};
        \draw[] (-1.5,1.4,0) node{$\mc{X}_B$};
      \end{scope}
    \end{tikzpicture}
  }
  \quad
  \subfloat[Histograms $f_A,f_B$ of $\mc{X}_A,\mc{X}_B$]{
    \begin{tikzpicture}
      \begin{axis}[
          width = 0.32\textwidth,
          height = 0.23\textwidth,
          axis lines = middle,
          xmin = -1,
          xmax = 11,
          ymin = -1,
          ymax = 7,
          ticks = none,
          x label style={anchor=north},
          y label style={anchor=west},
          xlabel = $\mc{A}$,
          ylabel = Probability,
        ]
        \addplot[domain=0:10.5,samples=100,color=black,thick,fill=green,fill opacity=0.2]{15*(x/3)*exp(-x^2/6)};
        \node at (axis cs:2.7,5.5) {$f_A$};
        \addplot[domain=0:10.5,samples=100,color=black,thick,fill=blue,fill opacity=0.2]{15*(x/10)*exp(-x^2/20)};
        \node at (axis cs:5,3.1) {$f_B$};
      \end{axis}
    \end{tikzpicture}
  }
  \caption{The lesion detectability problem is illustrated. (a) Given the image function $\phi$ from Fig. \ref{fig:bmode}b, we select two regions of interest (ROIs) $\mc{X}_A$ and $\mc{X}_B$. (b) The histograms of $\phi$ in $\mc{X}_A$ and $\mc{X}_B$ are obtained as $f_A$ and $f_B$, respectively.}
  \label{fig:lesdet}
\end{figure}
To define an image quality criterion, we need to restrict the statistical analysis of image values to a series of regions of interests (ROIs). Such regions typically correspond to the interior and neighborhood of a lesion when computing the CR, for example. The process of selecting ROIs amounts to choosing subsets of the full FOV of the image $\FOV_A,\FOV_B,\ldots\subset\FOV$. Throughout this work, we use uppercase letters (e.g., $A$, $B$) to denote the ROI image values, and subscripts on $\mc{X}$ (e.g., $\mc{X}_A$, $\mc{X}_B$) to refer to the spatial domains themselves. We denote the restriction of $\phi$ to an ROI $A$ as $\phi_{A}:\FOV_{A}\rightarrow\mc{A}$. An arbitrary ROI sub-image $A$ also possesses its own histogram $f_A:\mc{A}\rightarrow[0, \infty)$, verifying \eqref{eq:PDF_def} for $a=\phi_{A}(x)$.

Let us express the image values of two ROIs via the PDFs $f_A$ and $f_B$, respectively. Fig.~\ref{fig:lesdet} illustrates this scenario. As before, we can compute the image statistics within $A$ and $B$. With a slight abuse of notation, the means of  image values within the ROIs are then
\begin{align}
  \mbb{E}[A] & = \int_{\mc{A}} a\,f_A(a)\dd a,\
  \mbb{E}[B] = \int_{\mc{A}} b\,f_B(b)\dd b,
\end{align}
whereas the variances are defined as
\begin{align}
  \Var[A] & = \mbb{E}[A^2] - (\mbb{E}[A])^2,
  \nonumber
  \\
  \Var[B] &= \mbb{E}[B^2] - (\mbb{E}[B])^2.
\end{align}

\subsection{Image Quality Criteria}
The mean and variance are the main building blocks of many image quality criteria. An \emph{image quality criterion} is
defined as a function $q:\mc{A}^{\FOV}\rightarrow \mbb{R}$ that assigns a quality score $q(\phi)$ to any image $\phi\in\mc{A}^{\FOV}$, where $\mc{A}^{\FOV}$ is the set of all possible images. This is a general definition that includes many familiar image quality criteria that are used widely throughout medical imaging. When working with real-valued images, examples of $q(\phi)$ include:
\begin{align}
  \textrm{CR}(\phi;\FOV_A,\FOV_B)          & = \frac{\mbb{E}[A]}{\mbb{E}[B]}                                                                \\
  \textrm{CNR}(\phi; \FOV_A, \FOV_B)       & = \frac{\mbb{E}[A] - \mbb{E}[B]}{\sqrt{\Var[A]+\Var[B]}}                                       \\
  \textrm{SNR}(\phi; \FOV_A)               & = \frac{\mbb{E}[A]}{\sqrt{\Var[A]}}
  \\
  \textrm{FWHM}_\textrm{pt}(\phi;\FOV_A)   & = 2\argmin_{\Delta x} \left\{ \frac{\phi_{\!A}(x+\Delta x)}{\phi_{\!A}(x)} \ge 0.5 \right\}    \\
  \textrm{FWHM}_\textrm{corr}(\phi;\FOV_A) & = 2\argmin_{\Delta x} \left\{ \frac{R_{\phi\phi}(\Delta x)}{R_{\phi\phi}(0)} \ge 0.5 \right\}  \\
  \textrm{gCNR}(\phi;\FOV_A,\FOV_B)        & = 1 - \int_\mc{A} \min\{f_A(a), f_B(a)\}\dd a                                                  \\
  \textrm{FWHM}_\textrm{info}(\phi;\FOV_A) & = 2\argmin_{\Delta x} \left\{ \frac{I_{\phi\phi}(\Delta x)}{I_{\phi\phi}(0)} \ge 0.5 \right\},
\end{align}
where $R_{\phi\phi}(\Delta x) = \mbb{E}[\phi_A(x) \phi_A^*(x+\Delta x)]$ is the autocorrelation and $I_{\phi\phi}(\Delta x) = \mbb{E}[\log\frac{f_{AB}}{f_A f_B}]$ is the self-mutual information (``autoinformation'') of a translating ROI \cite{wagner1988fundamental,hyun2021info}. Moving forward, we use the shorthand $q[A,B]$ for a criterion $q(\phi;\FOV_A,\FOV_B)$.

Importantly, these criteria implicitly assume that image values are directly comparable across reconstruction methods, an assumption that becomes problematic when image values are altered by nonlinear transformations applied either during beamforming or post-processing.

\subsection{Dynamic Range Transformations}
Modern beamformers and image processing pipelines typically combine operations that alter the information content of the image (e.g., clutter suppression or adaptive weighting) with intensity remappings that primarily affect the dynamic range of image values. These dynamic range transformations (DRTs) may be applied explicitly for visualization or arise implicitly as part of nonlinear processing. As shown by Hverven et al. \cite{hverven2017influence} and Rindal et al. \cite{rindal2019effect}, such transformations can induce substantial changes in the statistical properties of image intensities and therefore cannot be ignored when evaluating image quality. We now formalize this discussion by introducing a precise definition of DRTs and their invariance classes.

\subsubsection{Dynamic range transformation (DRT)}
A DRT is a function $h:\mc{A}\rightarrow\mc{A}'$ that remaps the image values $\phi(x)\in\mc{A}$ into new values $h(\phi(x))\in\mc{A}'$. The new image values $\mc{A}'$ need not be the same as $\mc{A}$. As defined here, a DRT is a global operation that acts identically on all image position $x\in\FOV$. Let us denote the set of all DRTs as $\mc{T}_{\mathrm{all}}=\{h:\mc{A}\rightarrow\mc{A}'\}$. An image quality criterion $q$ is said to be \emph{invariant} under a class of DRTs $\mc{T}\subseteq\mc{T}_{\mathrm{all}}$ if
\begin{align}
  q(h(\phi)) = q(\phi) \quad \forall h\in\mc{T}.
\end{align}

\subsubsection{Invariance to scalar multiplication}
A \emph{(positive) scalar multiplication} is a function that multiplies the entire image $\phi$ by some real constant $c > 0$, i.e. $h(\phi(x)) = c\,\phi(x)$. Denote the set of all such functions as $\mc{T}_\times$. These DRTs are often taken for granted (e.g., normalizing an image by its maximum value) because ultrasound echoes are usually considered to have arbitrary units. These DRTs are \emph{ratio-preserving}, i.e. $a/b = h(a)/h(b)$ for any $a,b\in\mc{A}$ and $b\ne 0$.

\subsubsection{Invariance to monotonic transformation}
A \emph{(strictly) monotonic transformation} is a DRT such that $h(a) < h(b)$ if and only if $a < b$. Denote the set of all such DRTs as $\mc{T}_<$. These DRTs are ubiquitous in high-dynamic range imaging systems designed for human observers, e.g., grayscale compression. These DRTs are \emph{order-preserving}, but not necessarily ratio-preserving.

\subsubsection{Invariance to injective transformation}
An \emph{injective transformation} is a DRT such that every element $a\in\mc{A}$ maps to a unique element $a'\in\mc{A}'$. An injective DRT does not need to be monotonic. The set of all such DRTs is denoted $\mc{T}_\textrm{inj}$. A counterexample is quantization, where an interval of values in $\mc{A}$ map to the same value in $\mc{A}'$. Importantly, injective DRTs are \emph{information-preserving}: they have no impact on the entropy or mutual information of ROIs \cite{hyun2021ultrasound,hyun2021info}. Thus, injective DRTs can be thought of as DRTs that have no impact on an information-theoretic ideal observer.

These DRT classes form a strict hierarchy:
\begin{align}
  \mc{T}_{\mathrm{all}} \supset \mc{T}_\textrm{inj} \supset \mc{T}_< \supset \mc{T}_\times.
  \label{eq:DRT_classes}
\end{align}
Restricting the DRT invariance class preserves additional structure on the image value alphabet. For example, invariance to $\mc{T}_{\mathrm{inj}}$ preserves only distributional information, whereas invariance to $\mc{T}_<$ additionally preserves relative ordering of image values, and invariance to $\mc{T}_\times$ further retains scale and ratios.

This hierarchy is critical when evaluating nonlinear beamformers. For instance, the CR, CNR, SNR, FWHM$_\textrm{pt}$, and FWHM$_\textrm{corr}$ are $\mc{T}_\times$-invariant, whereas the gCNR and FWHM$_\textrm{info}$ are $\mc{T}_\textrm{inj}$-invariant, meaning that they can be applied rigorously to a much wider class of DRTs. However, transformations that are $\mc{T}_\textrm{inj}$-invariant but not $\mc{T}_<$-invariant are extremely unlikely to be used in a practical setting. In addition to changes in pixel ordering, they would necessarily imply discontinuous DRTs. Therefore, we argue that it is unnecessary to consider the full class of $\mc{T}_\textrm{inj}$-invariant functions when designing an imaging criterion, and that the analysis can be restricted to $\mc{T}_<$-invariant transformations. To the best of our knowledge, there are no $\mc{T}_<$-invariant criteria in use today, and we propose two such candidates below. 
\section{The Contrast Order Criterion}
In this section, we define the contrast order, establish its invariance properties, and relate it analytically to conventional contrast metrics under ideal speckle assumptions.
\label{sec:contrastorder}
\subsection{Definition of the Contrast Order}
\emph{Contrast} refers to the ability to distinguish two ROIs, e.g., lesion and background. We seek a new measure of contrast that describes the statistical ``orderability'' of the two ROIs, i.e., whether one ROI has values less or greater than the other, but without relying on the magnitude of the difference between them.

As before, let the random variables $A$ and $B$ correspond to the image values in ROIs $\mc{X}_A$ and $\mc{X}_B$ in $\mc{X}$ with probability distributions $f_A$ and $f_B$. We introduce a new criterion called the \emph{contrast order (CO)}, defined as
\begin{align}
  \textrm{CO}[A,B] & = \mbb{E}[\sign (A - B)]                                 \\
                   & = \iint \sign(a-b)\,f_{A}(a)\,f_{B}(b) \dd a\dd b, \label{eq:CO}
\end{align}
where the sign function is defined for any scalar $c\in\mbb{R}$ as
\begin{align}
  \sign (c) = \begin{dcases}
                +1, & c > 0 \\
                0,  & c=0   \\
                -1, & c < 0
              \end{dcases}.
\end{align}
This definition emphasizes relative ordering rather than magnitude. Just as the contrast ratio measures the \emph{ratio} between two ROIs, the contrast order measures the relative \emph{order} of the values.

\subsection{Properties of the Contrast Order}
The contrast order is bounded in the interval $[-1,+1]$, and is equal to $-1$ when all values in $\mc{X}_A$ are smaller than all values in $\mc{X}_B$ and is equal to $+1$ when all the values are larger. The contrast order is also an \emph{antisymmetric} function, i.e. $\textrm{CO}[A,B]=-\textrm{CO}[B,A]$, since $\sign(a-b)=-\sign(b-a)$ for all $a,b\in\mbb{R}$.

Now define a strictly monotonic transformation as a function $h:\mc{A}\rightarrow\mc{A}$ such that $h(a) < h(b)$ if and only if $a < b$.
\begin{theorem}[Invariance]
  \label{th:invariance}
  The contrast order is invariant under all strictly monotonic transformations.
  \proof
  First, observe that strictly monotonic transformations preserve the sign of the difference $a-b$ for all $a,b\in\mbb{R}$ by directly examining all possible cases of the sign function:
  \begin{align}
    \sign(h(a) - h(b)) & = \begin{dcases}
                             +1, & h(a)>h(b) \Longleftrightarrow a > b \\
                             0,  & h(a)=h(b) \Longleftrightarrow a = b \\
                             -1, & h(a)<h(b) \Longleftrightarrow a < b
                           \end{dcases} \nonumber \\
                       & = \sign(a - b),
  \end{align}
  which follows directly from the definition of a strictly monotonic transformation. Thus,
  \begin{align}
    \textrm{CO}[h(A),h(B)] & = \mbb{E}[\sign (h(A) - h(B))] \\
                           & = \mbb{E}[\sign (A-B)]         \\
                           & = \textrm{CO}[A,B],
  \end{align}
  i.e., the contrast order between two image ROIs is the same after composition with a strictly monotonic transformation.\qed
\end{theorem}
Theorem \ref{th:invariance} shows that the contrast order is preserved under operations like power compression $h(a)=a^p$ and logarithmic compression $h(a)=\log a$, both of which are commonly used in post-processing. Theorem~\ref{th:invariance} does not hold for general non-strict monotonic transformations, defined as functions $g:\mc{A}\rightarrow\mc{A}$ such that $g(a)\le g(b)$ if and only if $a \le b$.
\begin{corollary}
  \label{th:generalmonotonic}
  The contrast order of $g(A)$ and $g(B)$ may differ in magnitude and sign from the contrast order of $A$ and $B$ for general (non-strict) monotonic transformations $g$.
  \proof
  We use a simple counterexample. Consider an alphabet $\mc{A}=\{1,2,3\}$ and monotonic transformations $g_1(\mc{A})=\{1,2,2\}$ and $g_2(\mc{A})=\{1,1,2\}$. Let $A$ and $B$ be distributed as $f_A = \{0, 1, 0\}$ and $f_B = \{0.5, 0, 0.5\}$. The contrast order is computed via \eqref{eq:CO} to be:
  \begin{itemize}
    \item $\textrm{CO}[A,B] = 0$
    \item $\textrm{CO}[g_1(A),g_1(B)] = +0.5$
    \item $\textrm{CO}[g_2(A),g_2(B)] = -0.5$
  \end{itemize}
  Therefore, the magnitude and sign are not preserved under general monotonic transformations. \qed
\end{corollary}
This counterexample is intentionally extreme, and such large variations are unlikely to occur in practice. Nevertheless, this fact illustrates that strong CO variations can occur when the pixel intensities are concentrated around specific values or when non-strict monotonic transformations are employed, thus inviting care in interpreting the CO in such situations. A realistic example is amplitude quantization, where intervals of values are represented by individual values, e.g., when displaying an image on a monitor, or during histogram binning \cite{hyun2021ultrasound}.

\begin{proposition}
  \label{th:prob}
  {The contrast order is the probability of $A>B$ minus the probability of $A<B$.}
  \proof{
    Observe that for any $a,b\in\mc{A}\subset\mbb{R}$,
    \begin{align}
      \sign(a - b) = \mb{1}\{a > b\} - \mb{1}\{a < b\},
    \end{align}
    where $\mb{1}\{\cdot\}$ is the indicator function that is equal to 1 when the predicate $\{\cdot\}$ is true and $0$ otherwise. Also observe that
    \begin{align}
      \mbb{E}[\mb{1}\{A > B\}]
       & = \textrm{Pr}[A > B].
    \end{align}
    Then we have that
    \begin{align}
      \textrm{CO}[A,B] & = \mbb{E}[\sign (A - B)] \nonumber                                            \\
                       & = \mbb{E}[\mb{1}\{A > B\}] - \mbb{E}[\mb{1}\{A < B\}] \label{eq:co_indicator} \\
                       & = \textrm{Pr}[A > B] - \textrm{Pr}[A < B].
    \end{align}
    giving a probabilistic interpretation of the contrast order.}\qed
\end{proposition}

\subsection{The Contrast Order of Speckle Amplitude}
Speckle is caused by diffuse sub-resolution scattering. Consider a ROI $A$ presenting homogeneous fully-developed speckle with \emph{echogenicity} $\sigma_A$. Its amplitude within $\mc{X}_A$ follows a Rayleigh distribution with scale parameter $\sigma_A$:
\begin{align}
  f_A(a;\sigma_A) = \frac{a}{\sigma_A^2}\exp\left[-\frac{a^2}{2\sigma_A^2}\right],\qquad a\ge0.
\end{align}
The cumulative density function of $A$ is
\begin{align}
  F_A(a;\sigma_A) = \int_0^a f_A(a;\sigma_A)\dd a= 1 - \exp\left[-\frac{a^2}{2\sigma_A^2}\right].
\end{align}

We consider now two ROIs $A$ and $B$ with fully-developed speckle and echogenicities $\sigma_A$ and $\sigma_B$, respectively.
% Consider two speckle random variables $A$ and $B$ distributed as $f_A(a;\sigma_A)$ and $f_B(b;\sigma_B)$.
Their echogenicity ratio is defined as
\begin{align}
  \gamma = \frac{\sigma_A}{\sigma_B}.
\end{align}
We omit the scale parameters in the notation below for brevity.
\begin{proposition}
  The CR between two speckle signals $A$ and $B$ is equal to the echogenicity ratio $\gamma$.
  \proof{
    Observe that
    \begin{align}
      \textrm{CR}[A,B] = \frac{\mbb{E}[A]}{\mbb{E}[B]} = \frac{\sigma_A\sqrt{\pi/2}}{\sigma_B\sqrt{\pi/2}}=\gamma,
    \end{align}
    where we used the fact that the mean value of a Rayleigh random variable is $\sigma\sqrt{\pi/2}$.
  }\qed
\end{proposition}

\begin{theorem}{The contrast order between two speckle signals $A$ and $B$ is $\frac{\gamma^2-1}{\gamma^2+1}$.}
  \label{th:speckleco}
  \proof{
    We begin with \eqref{eq:co_indicator} from Proposition~\ref{th:prob}:
    \begin{align}
      \textrm{CO}[A,B] =& \mbb{E}[\mb{1}\{A > B\}] - \mbb{E}[\mb{1}\{A < B\}]
      \nonumber                                                                                                 \\
                       =& \iint\limits_{a>b} f_{A}(a)f_B(b)\dd a\dd b -
                       \nonumber
                       \\
                       & \iint\limits_{a<b} f_{A}(a) f_{B}(b)\dd a\dd b.
    \end{align}
    The first integral can be simplified as
    \begin{align}
      &\int\limits_0^\infty f_{A}(a) \left(\int\limits_0^{a} f_{B}(b)\dd b\right)\dd a                                                                                                   \\
                         =& \int_0^\infty f_{A}(a) \,F_{B}(a)\, \dd a \label{eq:co_1a}                                                                                                   \\
                         =& \int_0^\infty \frac{a}{\sigma_A^2}\exp\left[-\frac{a^2}{2\sigma_A^2}\right] \left(1-\exp\left[-\frac{a^2}{2\sigma_B^2}\right]\right) \dd a  \label{eq:co_1b} \\
                         =& 1 - \int_0^\infty \frac{a}{\sigma_A^2}\exp\left[-\frac{a^2(\sigma_A^2+\sigma_B^2)}{2\sigma_A^2\sigma_B^2}\right] \dd a  \label{eq:co_1c}                     \\
                         =& 1 - \frac{\sigma_B^2}{\sigma_A^2+\sigma_B^2}                                                                                                                 \\
                         =& 1 - \frac{1}{\gamma^2+1}.
    \end{align}
    The second integral is computed similarly as
    \begin{align}
      &\int\limits_0^\infty f_{A}(a) \left(\int\limits_a^{\infty} f_{B}(b)\dd b\right)\dd a \\
                         =& \int_0^\infty f_{A}(a) \left(1-F_{B}(a)\right) \dd a            \\
                         =& \frac{1}{\gamma^2+1}.
    \end{align}
    Thus the contrast order is
    \begin{align}
      \textrm{CO}[A,B] & = 1 - \frac{2}{\gamma^2+1} = \frac{\gamma^2-1}{\gamma^2+1},
      \label{eq:co_cr}
    \end{align}
    completing the proof.}\qed
\end{theorem}

\subsection{The Contrast Order of Speckle Intensity}
Now consider speckle \emph{intensities}, i.e. $A^2$ and $B^2$. Note that the squaring function $h(a) = a^2$ is strictly monotonic on the domain $a \ge 0$. Therefore, we expect the contrast order to be invariant under squaring ($\textrm{CO}[A^2,B^2]=\textrm{CO}[A,B]$), but not the CR ($\textrm{CR}[A^2,B^2]\ne\textrm{CR}[A,B]$). Let us prove these.

We will make use of the fact that $A^2$ is exponentially-distributed when $A$ is Rayleigh-distributed, with distribution
\begin{align}
  f_{A^2}(a';\sigma) = \frac{1}{2\sigma^2}\exp\left[-\frac{a'}{2\sigma^2}\right],\qquad a'>0
\end{align}
and cumulative distribution
\begin{align}
  F_{A^2}(a';\sigma) = \int_0^a f_A(a';\sigma)\dd a'= 1 - \exp\left[-\frac{a'}{2\sigma^2}\right].
\end{align}
\begin{proposition}
  The CR between two speckle intensities $A^2$ and $B^2$ is $\gamma^2$.
  \proof{
    The expected value of an exponential variable is $2\sigma^2$:
    \begin{align}
      \textrm{CR}[A^2,B^2] = \frac{\mbb{E}[A^2]}{\mbb{E}[B^2]} = \frac{2\sigma_A^2}{2\sigma_B^2}=\gamma^2,
    \end{align}
  }
  which is the square of the CR of $A$ and $B$.\qed
\end{proposition}

\begin{theorem}{The contrast order between two speckle intensities $A^2$ and $B^2$ is $\frac{\gamma^2-1}{\gamma^2+1}$.}
  \proof{
    Using the same procedure as Theorem \ref{th:speckleco}, we have
    \begin{align}
      \textrm{CO} [A^2,B^2]=&
      \mbb{E}[\mb{1}\{A^2 > B^2\}] - \mbb{E}[\mb{1}\{A^2 < B^2\}]
      \nonumber                \\=& \iint\limits_{a'>b'} f_{A^2}(a') f_{B^2}(b') \dd a'\dd b' -
      \nonumber
      \\ 
      &\iint\limits_{a'<b'} f_{A^2}(a') f_{B^2}(b')\dd a'\dd b'.
    \end{align}
    As before, the first integral can be simplified as
    \begin{align}
      &\int\limits_0^\infty f_{A^2}(a') \left(\int\limits_0^{a'} f_{B^2}(b')\dd b'\right)\dd a' \\
                           =& \int_0^\infty f_{A^2}(a') \,F_{B^2}(a')\, \dd a'                  \\
                           =& 1 - \frac{1}{\gamma^2+1},
    \end{align}
    and the second integral as
    \begin{align}
      &\int\limits_0^\infty f_{A^2}(a') \left(\int\limits_a^{\infty} f_{B^2}(b')\dd b'\right)da' \\
                           =& \int_0^\infty f_{A^2}(a') \left(1-F_{B^2}(a')\right) \dd a'        \\
                           =& \frac{1}{\gamma^2+1}.
    \end{align}
    Thus the contrast order is once again
    \begin{align}
      \textrm{CO}[A^2,B^2] & = 1 - \frac{2}{\gamma^2+1} = \frac{\gamma^2-1}{\gamma^2+1},
    \end{align}
    completing the proof.}\qed
\end{theorem}
Extrapolating further, Theorem \ref{th:invariance} mathematically guarantees that the contrast order of any strictly monotonic transformation $h$ of speckle signals $A$ and $B$ will always be
\begin{align}
  \textrm{CO}[h(A), h(B)] = \frac{\gamma^2 - 1}{\gamma^2 + 1}.
\end{align}

\subsection{Effective Contrast Ratio}
When the images do not follow traditional speckle statistics \cite{hverven2017influence,rindal2019effect}, the CR becomes decoupled from the echogenicity ratio $\gamma$, and its interpretation becomes unclear. Instead, we propose the \emph{effective contrast ratio (ECR)}, obtained by inverting \eqref{eq:co_cr}:
\begin{align}
  \textrm{ECR}[A,B] & = \sqrt{\frac{1+\textrm{CO}[A,B]}{1-\textrm{CO}[A,B]}}.
\end{align}
In words, ECR is the echogenicity ratio that two ideal speckle ROIs must have in order to reproduce a given contrast order value. Although the ECR is less fundamental than the contrast order itself, it provides a convenient calibration of the measured contrast order with the familiar concept of Rayleigh speckle contrast. When expressed in a logarithmic scale, it can also ease comparison between beamformers when lesions with high contrasts are considered by circumventing the saturation of the CO. We postulate that the ECR serves the intended purpose of current CR measurements (to measure the effective separability and relative ordering of two ROIs), while additionally being $\mc{T}_<$-invariant. Moving forward, we suggest that ECR should be used as a drop-in replacement for CR when comparing nonlinear beamformers.

\subsection{Estimating the Contrast Order}
\label{subsec:co_est}
In practice, the contrast order must be estimated from a finite number of image samples within each ROI. Let $\{A_i\}_{i=1}^{N_A}\sim A$ and $\{B_j\}_{j=1}^{N_B}\sim B$ denote samples drawn from ROIs $\mc{X}_A$ and $\mc{X}_B$, respectively. The natural estimator of the contrast order defined in \eqref{eq:CO} is obtained by averaging the sign of the difference over all cross-pairs:
\begin{align}
  \widehat{\textrm{CO}}[A,B] = \frac{1}{N_A N_B} \sum_{i=1}^{N_A}\sum_{j=1}^{N_B} \sign(A_i - B_j).
  \label{eq:CO_est}
\end{align}

Provided that the samples within each ROI are independent and identically distributed and that the two collections are mutually independent, this estimator is unbiased, i.e.,
\begin{align}
  \mathbb{E}[\widehat{\textrm{CO}}] = \textrm{CO}.
\end{align}
Furthermore, its variance has a simple and interpretable bound:
\begin{align}
  \label{eq:var_bound}
  \textrm{Var}[\widehat{\textrm{CO}}] \le \frac{1}{N_A} + \frac{1}{N_B} - \frac{1}{N_AN_B}.
\end{align}
Proofs of these results are provided in Appendix~\ref{sec:estimator}, whereas Appendix~\ref{sec:code} presents example MATLAB and Python implementations.

\begin{figure}
    \centering
    \includegraphics[]{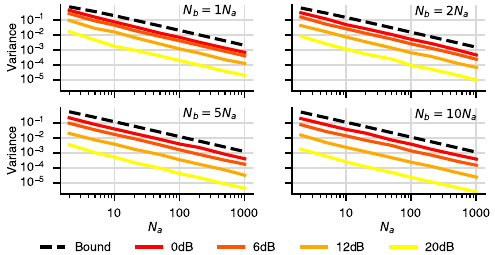}
    \caption{Comparison of the theoretical variance bound \eqref{eq:var_bound} with variance values obtained from synthetic samples generated from pairs of Rayleigh distributions with different contrast ratios.}
    \label{fig:bound}
\end{figure}

Equation~\eqref{eq:var_bound} provides an upper bound to the variance of the CO estimator. To analyze how tight the bound is in practice, we generate pairs of synthetic signals with sizes $N_a$ and $N_b$ drawn from two Rayleigh distributions with 0, 6, 12, or 20 dB intrinsic CR. We then compute the variance of the CO estimator between the two signals over 1000 realizations of the experiment. The results are depicted in Fig.~\ref{fig:bound} for different values of $N_a$ and $N_b$. We see that the tightness of the bound decreases as the CR increases and that the minimum ratio between the measured variance and the bound---occurring when signals drawn from identical distributions are compared---is approximately a factor one half for the Rayleigh distribution considered here.

Importantly, although $\widehat{\textrm{CO}}$ averages $N_AN_B$ pairwise comparisons, these comparisons are not mutually independent, as many share common samples. Consequently, the dominant variance terms scale as $1/N_A+1/N_B$, rather than $1/(N_AN_B)$, reflecting the effective number of independent observations contributed by each ROI. In particular, both CR and CNR (among other metrics) also exhibit an inverse linear relationship between estimator variance and the number of samples. Therefore, the performance of the proposed CO estimator is expected to scale similarly with the size of the ROIs.

In practice, ROIs are often sampled more finely than the system resolution, resulting in spatially correlated image values. In this case, the estimator remains unbiased, but the variance decreases more slowly due to a reduced effective sample size, and the variance bound should instead be interpreted with respect to the effective number of independent samples rather than the raw pixel counts.

Finally, the simplicity of the CO estimator makes it straightforward to analyze and interpret statistically.
In particular, $\widehat{\textrm{CO}}$ is tuning-free, exactly unbiased under standard assumptions, and admits a closed-form variance bound, making it a particularly transparent and robust estimator for ROI separability. In contrast, the density estimation required to compute gCNR complicates considerably any statistical analysis of its estimator from a theoretical point of view, even though it has been shown empirically to be unbiased provided a rigorous implementation \cite{arnestad2024robust}.

\section{Examples}
\label{sec:examples}
\begin{figure*}[!p]
    \centering
    \subfloat[Simulated contrast lesions]{\includegraphics[width=1.\textwidth]{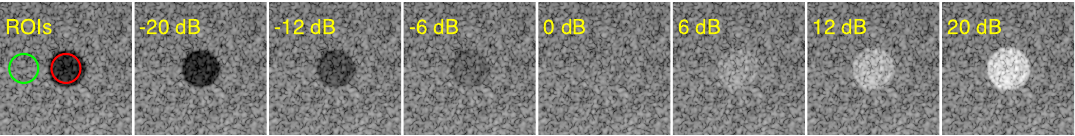}} \\
    \subfloat[Anechoic lesion with Gaussian noise added on received signals]{\includegraphics[width=1.\textwidth]{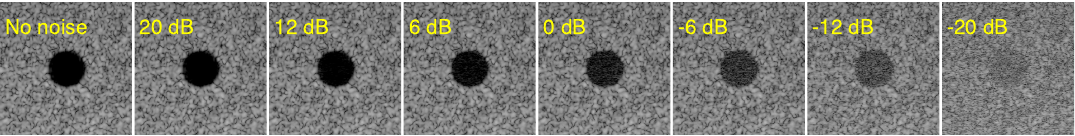}} \\
    \subfloat[Power compression of $-6$\,dB lesion]{\includegraphics[width=1.\textwidth]{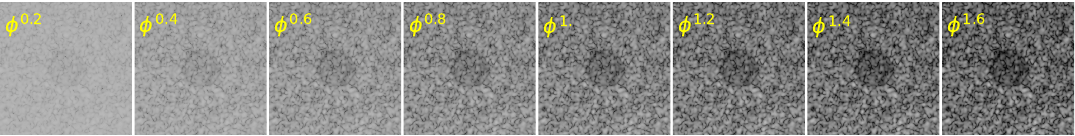}} \\
    \subfloat[$-6$\,dB lesion with increasing scatterer density]{\includegraphics[width=1.\textwidth]{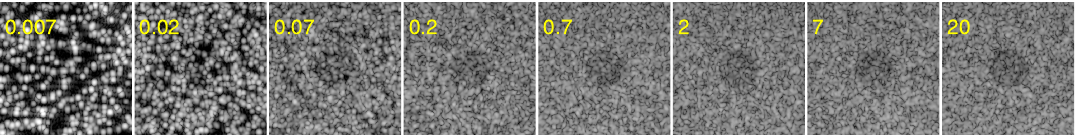}} \\[-0.05cm]
    \subfloat[Criteria vs. intrinsic contrast]{\includegraphics[width=.49\textwidth]{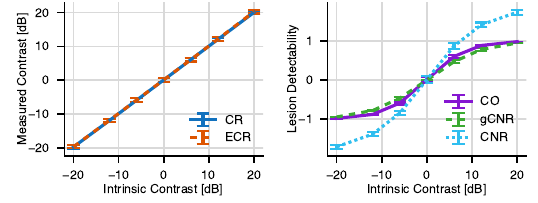}} \hfill
    \subfloat[Criteria vs. noise]{\includegraphics[width=.5\textwidth]{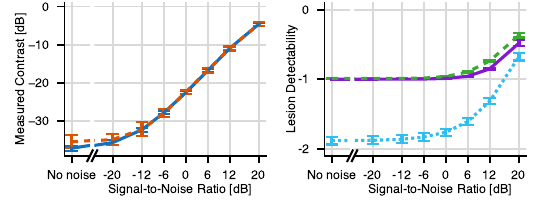}}\\[-0.05cm]
    \subfloat[Criteria vs. power compression]{\includegraphics[width=.5\textwidth]{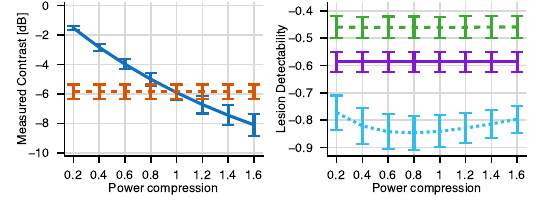}} \hfill
    \subfloat[Criteria vs. scatterer density]{\includegraphics[width=.5\textwidth]{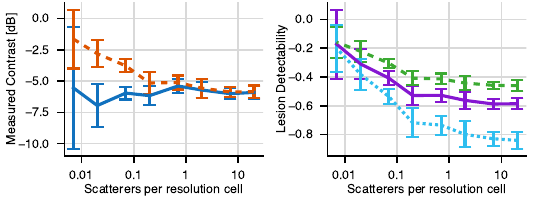}}
    \caption{Contrast lesions are simulated using Field II. (a) Lesions with increasing intrinsic echogenicity. (b) Anechoic lesion with Gaussian noise of increasing SNR added to the received signals. (c) ``Cosmetic'' changes in lesion contrast simulated using power compression for the -6 dB inclusion. (d) Changes in pixel value distribution obtained by varying the number of scatterers per resolution cells for the -6 dB inclusion.
    For the four cases, the CR, ECR, CO, (signed) gCNR, and (signed) CNR are plotted. (e) The CR and ECR behave identically when measuring true changes in lesion contrast. The CO varies from -1 to +1, as does the gCNR, and the CNR follows a similar trend. (f) All the metrics show similar behaviours with an increase in Gaussian noise.
    (g) The ECR of a $-6$\,dB lesion is invariant under power compression, whereas the CR changes. Similarly, the CO and gCNR are invariant, whereas the CNR varies. (h) The ECR, as well as the gCNR, SNR, and CO, show a strong decrease in lesion detectability when the number of scatterers is lowered, unlike the CR.}
	\label{fig:contrast_order}
\end{figure*}

\subsection{Contrast Order and ECR Demonstration}
\subsubsection{Datasets}
Field II simulations of lesion targets are used to demonstrate the behavior of the contrast order and ECR and compare them to the reference metrics. We rely on the same setting as the one proposed in \cite{hyun2016efficient}: a simulated L12-3v transducer (128 elements, 8\,MHz center frequency) imaging a speckle target with a full (multistatic) synthetic aperture sequence. The standard DAS beamformer is used in all cases with an f-number of 1 in transmission and reception.

Four different experiments are performed with simulated data. First, a 3\,mm-diameter cylindrical lesion is simulated at the elevation focus with 20 scatterers per resolution cell. Several values of the lesion's true echogenicity are tested, namely $-20$\,dB, $-12$\,dB, $-6$\,dB, $0$\,dB, $+6$\,dB, $+12$\,dB, and $+20$\,dB. The associated B-mode images are plotted in Fig.~\ref{fig:contrast_order}a. Second, we consider an anechoic lesion in the same position with the same number of scatterers per resolution cells. We study the noiseless case, as well as noisy cases where the acquired signals are corrupted by white Gaussian noises with signal-to-noise ratios (SNRs) of $20$\,dB, $12$\,dB, $6$\,dB, $0$\,dB, $-6$\,dB, $-12$\,dB, and $-20$\,dB. The images are depicted in Fig.~\ref{fig:contrast_order}b. Third, the $-6$\,dB lesion of the first experiment is studied under power compression by factors ranging from $0.2$ to $1.6$ to analyze the impact of DRTs on the different metrics. We display the images in Fig.~\ref{fig:contrast_order}c. Finally, we assess the proposed metrics with different reflectivity distributions, namely we consider a $-6$\,dB hypoechoic lesion with 8 different numbers of scatterers per resolution cell, $0.007$, $0.02$, $0.07$, $0.2$, $0.7$, $2$, $7$, and $20$. The B-mode images are plotted in Fig.~\ref{fig:contrast_order}d.

The lesion and background ROIs are depicted in the first panel of Fig.~\ref{fig:contrast_order}a. The CR, ECR, CO, gCNR, and CNR are computed for each experiment. We report signed versions of the gCNR and CNR by multiplying the sign of the mean difference to enable direct comparison with the antisymmetric contrast order. A total of 8 random scatterer realizations are simulated to quantify the standard deviations of the estimators.

\subsubsection{Results}
In Fig.~\ref{fig:contrast_order}e, the CR increases linearly with the true lesion contrast, as expected of the DAS beamformer. Likewise, the ECR also matches the CR precisely due to the Rayleigh statistics of the speckle. The CO, signed gCNR, and signed CNR follow the same sigmoidal shape, with diminishing returns in lesion detectability at higher contrast magnitude. As it can be seen in Fig.~\ref{fig:contrast_order}f similar conclusions can be drawn from the experiments with the anechoic lesion. ECR and CR coincide except at very low noise levels, where the image signal within the lesion is dominated by sidelobe scattering. The CO, gCNR, and CNR also quickly saturate as the SNR decreases. In Fig.~\ref{fig:contrast_order}g, we can observe that power compression affects the CR despite the true lesion contrast remaining the same at $-6$\,dB, while the ECR is unaffected. Similarly, the contrast order and gCNR are invariant under power compression, whereas the CNR changes values. Finally, in Fig.~\ref{fig:contrast_order}h, the CR remains approximately constant with the number of scatterers, despite a significant increase in its standard deviation as the number of scatterers decreases. In contrast, the ECR (like the CO, gCNR, and CNR) is reduced in magnitude with lower amounts of scatterers, as the inclusion becomes increasingly difficult to distinguish in Fig.~\ref{fig:contrast_order}d.

\begin{figure}
    \centering
    \includegraphics{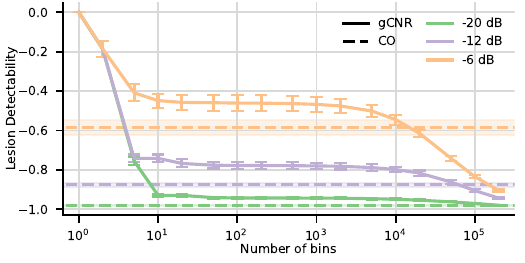}
    \caption{Dependence of gCNR on the number of histogram bins used for its computation when considering three inclusions at different contrast levels; CO is shown for comparison and does not depend on any tunable parameter.}
    \label{fig:var_bins}
\end{figure}

Fig.~\ref{fig:contrast_order}g is a numerical example of how the contrast order and ECR are invariant under a simple monotonic transformation, as proven in Theorem~\ref{th:invariance}. This invariance makes the ECR a better alternative to the CR for nonlinear beamformers. We also see that the ECR coincides with the CR for Rayleigh-distributed signals from a linear beamformer, and only deviates from CR when nonlinearities are introduced or when other echogenicity distributions are considered. Therefore, the ECR is a superior choice for measuring contrast differences independently of DRTs.

\subsubsection{Variability to Parameters}

In addition, we compare the contrast order and a naive implementation of gCNR with a specific focus on their dependence to tunable parameters. In all experiments performed up to now, we considered 64 histogram bins for the computation of the gCNR. However, changing the number of bins influences the gCNR, as we already highlighted in \cite{hyun2021ultrasound}. We illustrate this relationship in Fig.~\ref{fig:var_bins} for three hypoechoic inclusions with different intrinsic contrast, along with the contrast order, which is independent of any tunable parameter. We observe that gCNR is strongly biased if the number of bins is too large or too low, but admits a zone of stability, here between 20 and 1000 bins approximately. Nonetheless, a weak relationship to the number of bins still exists within this zone of stability (gCNR decreases from -0.457 to -0.468 for the -6 dB inclusion between 20 bins and 1000 bins), and the bounds of this zone of stability are influenced by the size of the ROIs and image spatial sampling. Even though more advanced implementations significantly mitigate the impact of tunable parameters \cite{arnestad2024robust}, this example illustrates that gCNR will always suffer from a fundamental ambiguity insofar as there are no unique natural ways of performing density estimation. There exists therefore a dependence on the implementation that may have an influence on the gCNR, especially when the amount of independent samples considered is low. In contrast, the CO admits, by construction, a canonical parameter-free estimator.

\subsection{Beamformer Comparisons}

\begin{figure*}[tb]
    \centering
    \includegraphics[width=.98\textwidth]{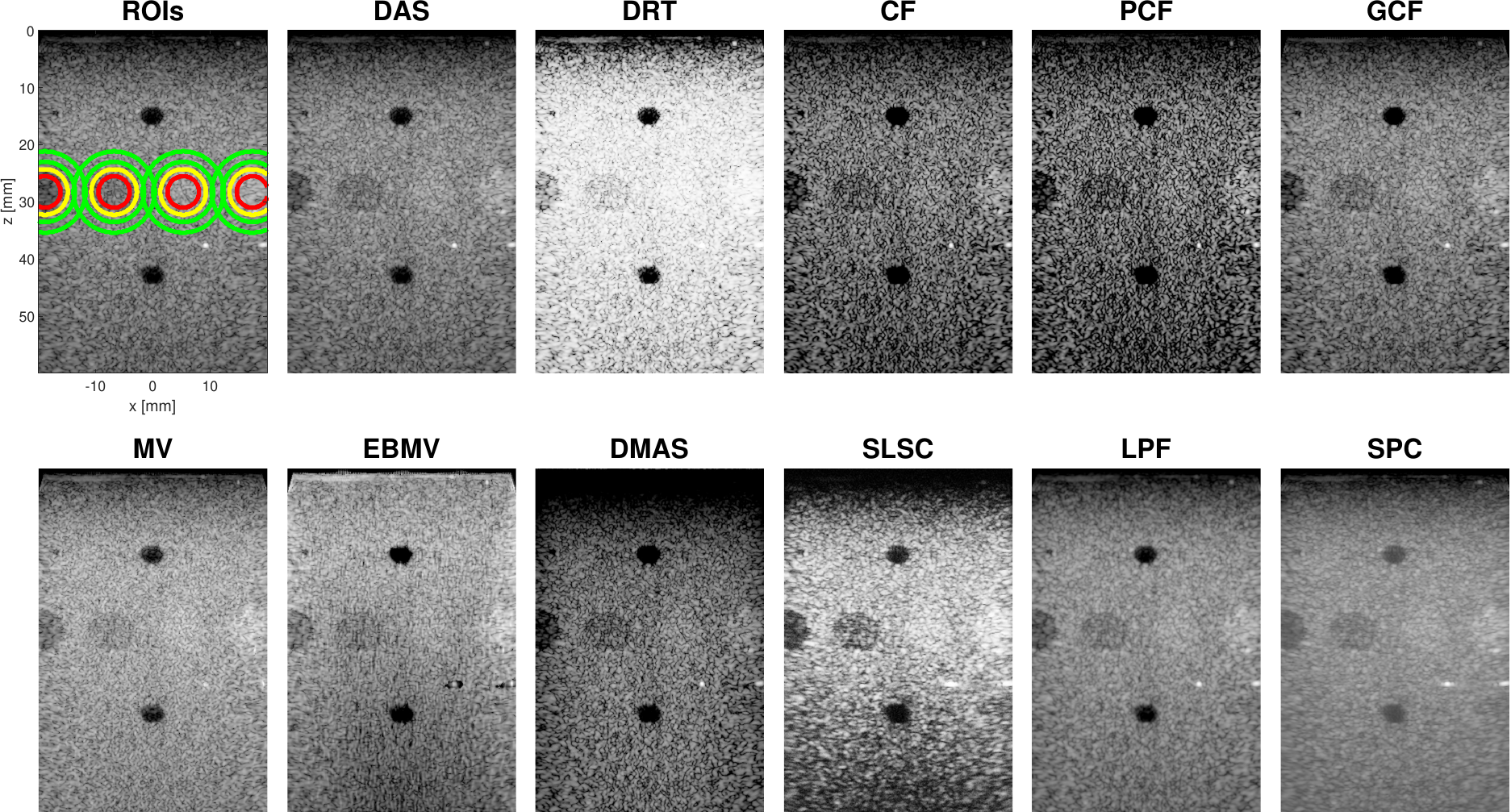}
    \caption{Multiple beamformers are used to reconstruct grayscale targets from the PICMUS experimental contrast phantom dataset \cite{liebgott2016plane}: delay-and-sum (DAS), a simple dynamic range transformation (DRT), coherence factor (CF), phase coherence factor (PCF), generalized coherence factor (GCF), Capon's minimum variance (MV), eigenspace-based minimum variance (EBMV), delay-multiply-and-sum (DMAS), short-lag spatial coherence (SLSC), a simple low-pass filter (LPF), and receive spatial compounding (SPC). The red inner circle denotes the lesion ROI, the yellow circle the lesion location, and green ring the background ROI. Note the wide variability in the image contrasts and textures.}
	\label{fig:picmus_images}
\end{figure*}

\begin{figure*}[tb]
    \centering
    \includegraphics[width=.98\textwidth]{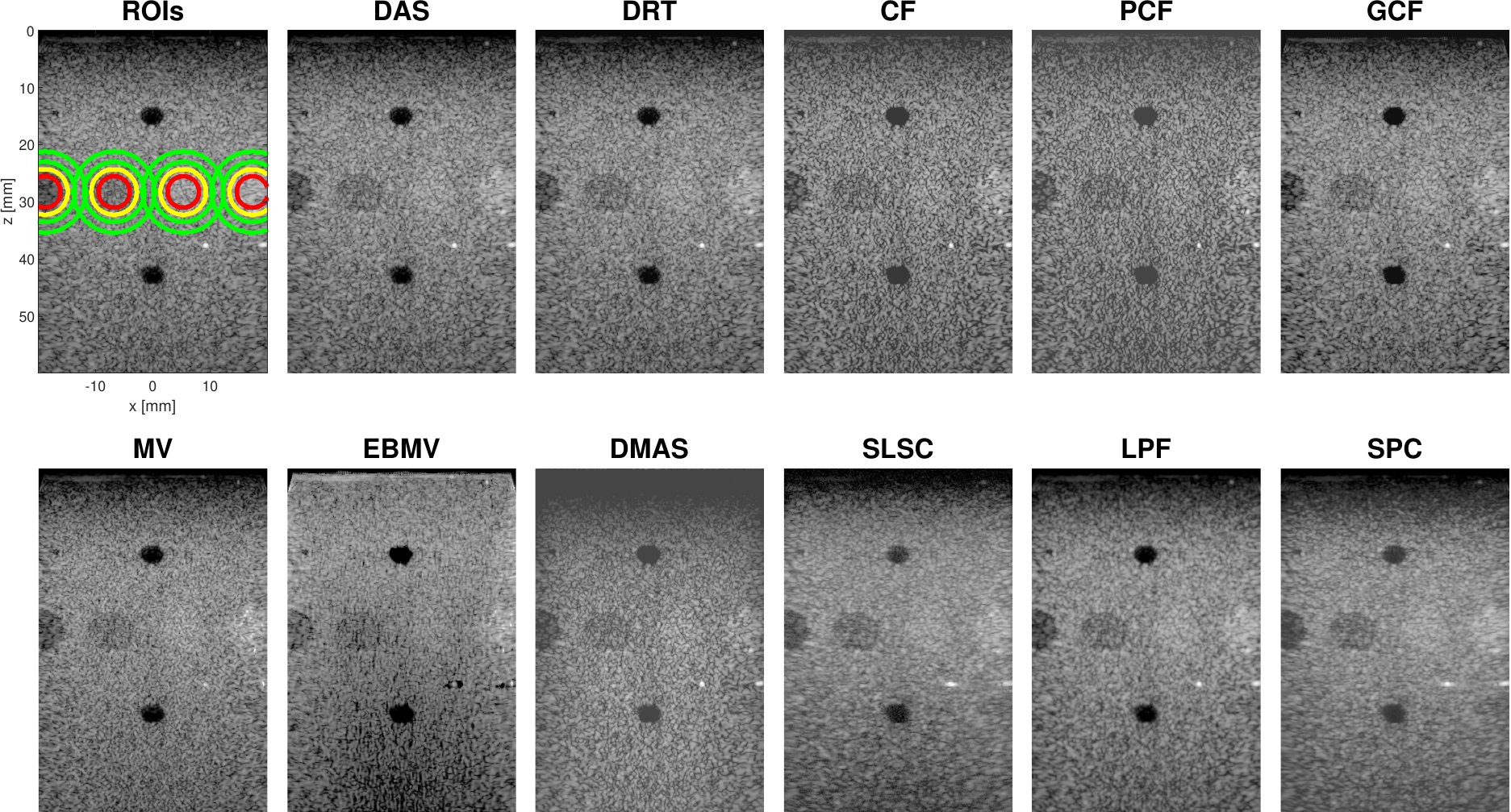}
    \caption{Histogram matching was applied to the beamformed images from Fig.~\ref{fig:picmus_images} using the DAS histogram as the reference. The resulting image textures and contrasts are visually more similar to one another, making them easier to compare.}
	\label{fig:picmus_hm_images}
\end{figure*}

% \begin{figure*}[tb]
%     \centering
%     \includegraphics[width=.98\textwidth]{images/picmus_quality}
%     \caption{The CR, ECR, (signed) CNR, and (signed) gCNR are plotted for each beamformer for each grayscale target in Fig.~\ref{fig:picmus_images}. Observe that the CR and CNR of the purely cosmetic DRT is different from DAS, showing their volatility. The CR and CNR, which depend on the image statistics, disagree with the trends of the ECR and gCNR, particularly for the $-6$\,dB lesion.}
% 	\label{fig:picmus_quality}
% \end{figure*}

\begin{figure*}[tb]
    \centering
    \includegraphics[width=\textwidth]{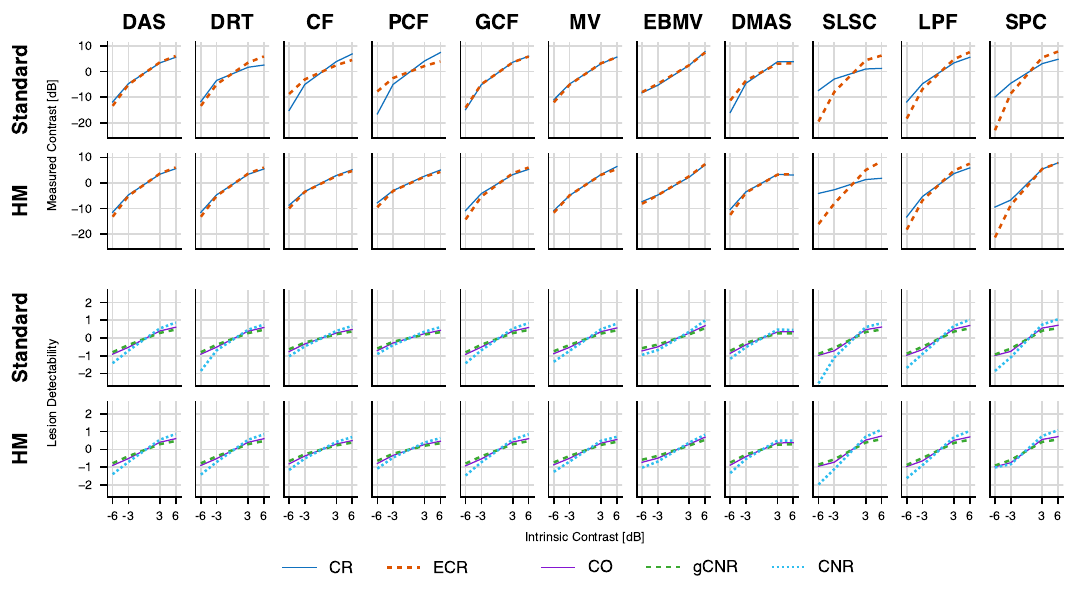}
    \caption{The CR and ECR (top rows), as well as the CO, (signed) gCNR, and (signed) CNR (bottom rows) plotted for each beamformer as a function of the nominal intrinsic contrast of the four greyscale targets. We compute the metrics using the images depicted in Fig.~\ref{fig:picmus_images} and Fig.~\ref{fig:picmus_hm_images}, before and after histogram matching (HM), respectively. Observe that the CR and CNR of the purely cosmetic DRT is different from DAS, showing their volatility. The CR and CNR, which depend on the image statistics, disagree with the trends of the ECR and gCNR, particularly for the $-6$\,dB lesion. HM affect the CR and CNR, notably removing the effects of the DRT. The ECR and gCNR remain unchanged.}
	\label{fig:picmus_quality}
\end{figure*}

This experiment evaluates how the contrast order and ECR vary across a diverse set of beamformers previously studied in the context of dynamic range sensitivity \cite{rindal2019effect}.
\subsubsection{Dataset}
A wide range of current beamformers were evaluated using the experimental contrast speckle dataset from the Plane-Wave Imaging Challenge in Medical Ultrasound (PICMUS) data \cite{liebgott2016plane}. This dataset consists of 75 plane wave transmissions, acquired with an L11-4v probe on a Verasonics Vantage 256 system. The reader is referred to \cite{liebgott2016plane} for further details. The imaging target was a CIRS Model 040GSE phantom containing several anechoic regions, as well as four grayscale targets of nominal contrast -6\,dB, -3\,dB, +3\,dB, and +6\,dB. The testing fixtures were slightly modified so as to select the grayscale targets rather than the anechoic targets. Concentric ROIs were selected, with $x$ centers \{$-18.9$\,mm$, -6.8$\,mm$, 5.3$\,mm$, 17.4$\,mm\}, and all centered at $z=28.5$\,mm. An inner circular ROI and outer ring ROI were selected with a padding of 2 times the lateral resolution (circular radius: $2.75$\,mm; ring inner radius: $5.25$\,mm; ring outer radius: $7.11$\,mm).

\subsubsection{Beamformers}
Following the implementation from \cite{rindal2019effect}, the following beamformers were tested: DAS; a purely cosmetic DRT (gray-level transform \cite{rindal2019effect} with $\alpha=0.12$, $\beta=50$, $\epsilon=0.012$); DAS weighted by coherence factor (CF) \cite{mallart1994adaptive}, phase coherence factor (PCF) \cite{camacho2009phase}, and generalized coherence factor (GCF) \cite{li2003adaptive}; the Capon minimum variance (MV) \cite{synnevag2009benefits} and eigenspace-based minimum variance (EBMV) beamformers \cite{asl2010eigenspace}; the filtered delay-multiply-and-sum (DMAS) beamformer \cite{matrone2014delay}; the short-lag spatial coherence (SLSC) beamformer \cite{lediju2011short}; as well as a simple dynamic range transformation (DRT), Gaussian low pass filter (LPF) with $\sigma=\lambda/3$, and 4$\times$ receive spatial compounding (SPC). Specifically for the SLSC beamformer, a $-6$\,dB root-mean-square (RMS) noise was added to the receive channel data to avoid excessive coherence artifacts from the noiseless simulation environment \cite{dahl2011lesion}. Unless otherwise stated above, we utilized the Ultrasound Toolbox (USTB)  \cite{rodriguez2017ultrasound} implementation and parameters for each beamformer as described by \cite{rindal2019effect}. We used an f-number of 1.75 in reception and weighted spatially the images to compensate for variations of echogenicity caused by the overlaps between plane wave supports.

\subsubsection{Histogram Matching}

We perform the comparison between the beamformers twice. First, we consider the raw output of the beamformers without any normalization. Second, we repeated the analysis, this time with an additional histogram matching (HM) step to equalize the qualitative features of the images \cite{bottenus2020histogram}. Histogram matching reduces visual variability across beamformers by enforcing a common image value distribution, therefore it strongly alleviates differences in the lesion detectability metrics that may be caused by changes in the global distribution of pixel intensities. Depending on how the histogram bins are selected, matching may or may not be strictly monotonic. The log-compressed DAS image was used as the reference, and all histogram matching was performed on log-compressed images, with the exception of the SLSC beamformer, whose units are correlation coefficients. Whole-image matching was performed using the \texttt{imhistmatch} function in MATLAB.

\subsubsection{Results}

Fig.~\ref{fig:picmus_images} and Fig.~\ref{fig:picmus_hm_images} show the reconstruction from each beamformer before and after HM, respectively. In addition, Fig.~\ref{fig:picmus_quality} compares the evolution of CR, ECR, CO, gCNR, and CNR with the nominal intrinsic contrast of the targets, with and without HM. 

As seen in Fig.~\ref{fig:picmus_images}, the image value distributions vary widely when HM is not used. CF, PCF, and DMAS tend to darken the image and increase speckle grain. DRT, MV, EBMV, and SLSC increase the overall brightness of the image. SLSC, LPF, and SPC also appear to have slightly worse resolution. LPF and SPC have worse contrast, but reduce speckle grain. These differences epitomize the challenges that faced modern ultrasound image quality assessment before the introduction of the gCNR, i.e., prior to 2019. The purely cosmetic DRT beamformer increases CNR signficantly over DAS, as previously observed by \cite{rindal2019effect}. Perhaps more concerningly, the CR and CNR disagree on the \emph{ranking} of beamformer quality: as seen in Fig.~\ref{fig:picmus_quality}, the top 3 performers in CR in the $-6$\,dB lesion are PCF, DMAS, and CF, but the top 3 in CNR are SLSC, DRT, and SPC.

These observations highlight the sensitivity of CR and CNR to changes in image value distributions across beamformers. Because these criteria depend explicitly on image statistics such as mean and variance, differences in overall brightness, contrast scaling, or distribution shape can substantially influence their values, even when the underlying ordering of image intensities within the ROIs remains unchanged.

By contrast, the ECR, CO, and gCNR exhibit substantially more consistent behavior across beamformers in Fig.~\ref{fig:picmus_quality}. They produce rankings that are consistent up to statistical variations falling within the expected variance of the estimators, and that are less affected by global shifts in image statistics. This suggests that DRT-invariant criteria provide a more stable basis for comparing beamformers whose outputs differ in dynamic range or statistical structure, without requiring additional normalization or post hoc adjustment.

As shown in Fig.~\ref{fig:picmus_hm_images}, the resulting images are more similar in grayscale tone and overall brightness when HM is used, facilitating qualitative visual comparison. As seen in Fig.~\ref{fig:picmus_quality}, the CR and CNR curves now exhibit a more similar behavior to those produced by the ECR and gCNR. In particular, beamformers whose CR and CNR values were previously elevated due to dynamic range effects exhibit curves consistent with those obtained using DRT-invariant criteria.

This behavior suggests that histogram matching can partially mitigate the sensitivity of CR and CNR to differences in image statistics by approximately compensating for dynamic range transformations. In this sense, histogram matching may be viewed as a valuable preprocessing step that enables more meaningful use of traditional contrast metrics for qualitative assessment and visualization.

However, the effectiveness of histogram matching depends on implementation choices, such as histogram binning, reference selection, and whether matching is performed globally or within ROIs \cite{bottenus2020histogram}. Moreover, histogram matching does not guarantee invariance under subsequent DRTs. Histogram matching can be viewed as an attempt to retrofit DRT invariance onto metrics that lack it by design. By contrast, DRT-invariant criteria such as the ECR and gCNR provide a guaranteed and principled means of comparing beamformers without requiring normalization or additional assumptions.

\section{Discussion}
\label{sec:discussion}
Modern ultrasound beamformers increasingly aim to improve the \emph{information content} of an image, e.g., suppressing off-axis clutter or emphasizing coherent signal components, often through nonlinear processing. These operations routinely alter image value distributions, making traditional contrast criteria such as CR and CNR difficult to interpret. Because these criteria depend explicitly on summary statistics like mean and variance, they cannot reliably distinguish between genuine improvements in ROI separability and changes induced by dynamic range remapping. This limitation complicates rigorous comparison across modern beamformers.

As demonstrated in Section~\ref{sec:examples}, histogram matching can partially mitigate this issue by enforcing a common image value distribution before evaluation, thereby restoring consistency among classical contrast rankings. However, histogram matching functions as a \emph{normalization strategy} rather than an intrinsic property of the criterion itself. Its effectiveness depends on implementation choices, such as whether histogram bins are selected on a logarithmic scale or whether matching is performed with respect to the whole image versus specific ROIs \cite{bottenus2020histogram}. In contrast, defining image quality criteria that are intrinsically invariant under DRTs eliminates the need for normalization altogether, allowing beamformers to be compared directly modulo monotonic remappings of image values.

Within this framework, the contrast order provides a simple, order-based measure of ROI separability that depends only on the relative ordering of image values. Its invariance under strictly monotonic DRTs follows directly from its reliance on the sign of pairwise comparisons, regardless of magnitude. The contrast order is naturally signed, indicating which ROI is brighter on average, and has a simple, unbiased estimator with a closed-form variance upper bound. These properties make the contrast order a particularly transparent and statistically robust criterion for evaluating image quality. The ECR further maps the contrast order onto a familiar Rayleigh-calibrated scale, preserving continuity with historical CR-based literature while retaining DRT invariance.

The close agreement between the contrast order and gCNR across a wide range of beamformers is encouraging, because the gCNR is explicitly tied to the performance of an information-theoretic ideal observer \cite{hyun2021ultrasound}. This suggests that the contrast order captures a notion of ROI separability that is closely aligned with ideal observer detectability, despite being defined through a far simpler construction. Importantly, any distinctions between images that are detectable by the gCNR but not by the contrast order would necessarily arise from non-monotonic reorderings of image values, which specifically do not correspond to meaningful changes in imaging pipelines or display mappings. From this perspective, the contrast order can be viewed as a tractable, order-preserving surrogate for the ideal observer that avoids sensitivity to mathematically permissible but practically irrelevant DRTs.

The contrast does not exist in isolation. In Section~\ref{sec:examples}, the SLSC, Gaussian blur, and spatial compounding images achieved high ECR and gCNR values while exhibiting visibly reduced resolution, highlighting the inherent tradeoff between contrast and resolution; as we have previously shown \cite{hyun2021ultrasound}, blurring ROIs improves the separability of their respective histograms. This underscores that contrast and resolution provide critical context for one another, and must be presented and interpreted jointly. Unfortunately, current resolution criteria like the point target FWHM and speckle autocorrelation FWHM are not DRT-invariant \cite{rindal2020resolution,wagner1988fundamental,hyun2021info}. Further work is necessary to identify satisfactory DRT-invariant resolution criteria. While Sparrow's resolution criterion \cite{sparrow1916spectroscopic} and the autoinformation \cite{hyun2021info} are strong candidates, more development and characterization are needed to ensure that they apply to relevant imaging scenarios.

Finally, DRT-invariant criteria open the door to deeper, more complex questions. At what point is the gain in contrast worth the sacrifice in resolution? After optimizing the information content of the image in a DRT-agnostic fashion, how do we find the DRT that is optimal for human vision? How should image quality criteria evolve as imaging pipelines increasingly incorporate nonlinear processing and automated analysis?

\section{Conclusion}
We have introduced a new order-based contrast criterion called the contrast order. The contrast order measures the ``orderability'' of two ROIs. It is antisymmetric and bounded in the interval $[-1,+1]$. Moreover, the contrast order is invariant under all strictly monotonic transformations, although it may vary under non-strict monotonic transformations.
We showed that the contrast order can be used to derive an \emph{effective} contrast ratio as $\textrm{ECR}=\sqrt{(1+\textrm{CO})/(1-\textrm{CO})}$. The ECR provides a transformation-invariant criterion that coincides with the CR for Rayleigh-distributed speckle, making it a suitable drop-in replacement for the popular CR criterion when evaluating nonlinear beamformers.
\appendix

\refstepcounter{section}
\section*{Appendix \thesection}
\label{sec:estimator}

Continuing from Sec.~\ref{subsec:co_est}, let $\{A_i\}_{i=1}^{N_A}\sim A$ and $\{B_j\}_{j=1}^{N_B}\sim B$ denote independent and identically-distributed (i.i.d.) samples drawn from ROIs $\mc{X}_A$ and $\mc{X}_B$, respectively, where the two collections are mutually independent.

\begin{proposition}{$\widehat{\textrm{CO}}$ is an unbiased estimator of CO.}
  \proof{
    The expectation of the estimator is
    \begin{align}
      \mathbb{E}[\widehat{\textrm{CO}}[A, B]]
       & = \mathbb{E}\left[\frac{1}{N_A N_B} \sum_{i=1}^{N_A}\sum_{j=1}^{N_B}  \sign(A_i - B_j) \right] \\
       & = \frac{1}{N_A N_B} \sum_{i=1}^{N_A}\sum_{j=1}^{N_B} \mathbb{E}\left[ \sign(A_i - B_j)\right]  \\
       & = \frac{1}{N_A N_B} \sum_{i=1}^{N_A}\sum_{j=1}^{N_B} \mathbb{E}\left[ \sign(A - B)\right]      \\
       & = \mathbb{E}\left[ \sign(A - B)\right]                                                         \\
       & = \textrm{CO}[A, B],
    \end{align}
    which equals the population contrast order.
    \qed
  }
\end{proposition}

\begin{proposition}{$\textrm{Var}(\widehat{\textrm{CO}})$ is upper-bounded by $\frac{1}{N_A}+\frac{1}{N_B}-\frac{1}{N_AN_B}$.}
  \proof{
  Let $X_{ij} = \textrm{sign}(A_i - B_j)$. Then,
  \begin{align}
    \textrm{Var}(\widehat{\textrm{CO}})
     & = \textrm{Var}\left(\frac{1}{N_A N_B}\sum_{i=1}^{N_A}\sum_{j=1}^{N_B} X_{ij}\right) \\
     & = \frac{1}{N_A^2 N_B^2}\sum_{i,j}\sum_{i',j'} \textrm{Cov}(X_{ij},X_{i'j'}),
    \label{eq:co_var_expand}
  \end{align}
  i.e., the variance expands into a double sum of covariances between ordered pairs $(i,j)$ and $(i',j')$.
  We proceed by considering splitting the covariance terms into four subsets:
  \begin{enumerate}
    \item $\sum_{i\ne i',j\ne j'}\textrm{Cov}(X_{ij},X_{i'j'}) = 0$
    \item $\sum_{i=i',j=j'}\textrm{Cov}(X_{ij},X_{i'j'}) = N_A N_B\, \textrm{Var}(X_{11})$
    \item $\sum_{i=i',j\ne j'}\textrm{Cov}(X_{ij},X_{i'j'}) = N_A N_B (N_B-1)\, \textrm{Cov}(X_{11},X_{12})$
    \item $\sum_{i\ne i',j=j'}\textrm{Cov}(X_{ij},X_{i'j'}) = N_B N_A (N_A-1)\, \textrm{Cov}(X_{11},X_{21})$
  \end{enumerate}
  In case 1, $(A_i,B_j)$ is independent of $(A_{i'},B_{j'})$ by the i.i.d.\ assumptions, leading to zero covariance. Case 2 is simply the variance terms. In cases 3 and 4, one of the indices is shared, leading to nonzero covariance. Note that in cases 2-4, the (co)variance does not depend on the specific indices due to the i.i.d.\ assumptions.

  It remains to develop the bounds of $\textrm{Var}(X_{11})$, $\textrm{Cov}(X_{11},X_{12})$, and $\textrm{Cov}(X_{11},X_{21})$. Since $X_{ij}\in\{-1,0,+1\}$, we have $|X_{ij}|\le 1$ and therefore
  \begin{align}
    \textrm{Var}(X_{11}) \le \mathbb{E}[X_{11}^2] \le 1.
  \end{align}
  Let us condition $\textrm{Cov}(X_{11},X_{12})$ on $A_1$. Given $A_1$, the random variables
  $X_{11}=\textrm{sign}(A_1-B_1)$ and $X_{12}=\textrm{sign}(A_1-B_2)$ are independent. Therefore, by the law of total covariance,
  \begin{align}
    \textrm{Cov} & (X_{11},X_{12})        \nonumber                                        \\
    =\;          & \mathbb{E}\!\left[\textrm{Cov}(X_{11},X_{12}\mid A_1)\right]
    + \textrm{Cov}\!\left(\mathbb{E}[X_{11}\mid A_1],\mathbb{E}[X_{12}\mid A_1]\right)     \\
    =\;          & 0 + \textrm{Var}\!\left(\mathbb{E}[\textrm{sign}(A_1-B)\mid A_1]\right) \\
    \le\;        & 1,
  \end{align}
  where the last inequality follows from $\mathbb{E}_B[\textrm{sign}(a-B)]\in[-1,1]$, which implies $\textrm{Var}(\mathbb{E}_B[\textrm{sign}(a-B)]) \le 1$. An analogous argument conditioning on $B_1$ yields $\textrm{Cov}(X_{11},X_{21})\le 1$.

  Substituting these inequalities into the covariance subsets, \eqref{eq:co_var_expand} becomes bounded as
  \begin{align}
    \textrm{Var}(\widehat{\textrm{CO}})
     & \le \frac{
      N_A N_B
      + N_A N_B(N_B-1)
      + N_B N_A(N_A-1)}{N_A^2 N_B^2}
    \\
     & = \frac{1}{N_A}+\frac{1}{N_B}-\frac{1}{N_A N_B}.
  \end{align}
  This proves the estimator variance upper bound. \qed
  }
\end{proposition}

\refstepcounter{section}
\section*{Appendix \thesection}
\label{sec:code}
We provide two examples of estimating the contrast order from the image values in two ROIs \texttt{img1} and \texttt{img2}, as well as the ECR, with emphasis on clarity over efficiency.

\subsection{MATLAB}
\begin{lstlisting}[language=Matlab]
function co = contrast_order(img1, img2)
    num = 0;  % Initialize running sum
    for i = 1:numel(img1)      % Loop over img1
        for j = 1:numel(img2)  % Loop over img2
            num = num + sign(img1(i) - img2(j));
        end
    end
    % Normalize and return output
    co = num / numel(img1) / numel(img2);
end

function ecr = effective_contrast_ratio(co)
    ecr = sqrt((1 + co) ./ (1 - co));
end
\end{lstlisting}

\subsection{Python}
\begin{lstlisting}[language=Python]
import numpy as np

def contrast_order(img1, img2):
    num = 0.0  # Initialize running sum
    for a in img1.ravel():      # Loop over img1
        for b in img2.ravel():  # Loop over img2
            num += np.sign(a - b)
    # Normalize and return output
    return num / img1.size / img2.size

def effective_contrast_ratio(co):
    return np.sqrt((1 + co) / (1 - co))
\end{lstlisting}

\bibliographystyle{ACM-Reference-Format}
\bibliography{IEEEabrv,references}

\end{document}